\documentclass[a4paper,aps,prl,twocolumn,superscriptaddress]{revtex4}

\usepackage[pdftex]{graphicx}
\usepackage{latexsym,amsmath,verbatim}

\begin{document}

\begin{figure}[h]
\includegraphics*[width=\linewidth]{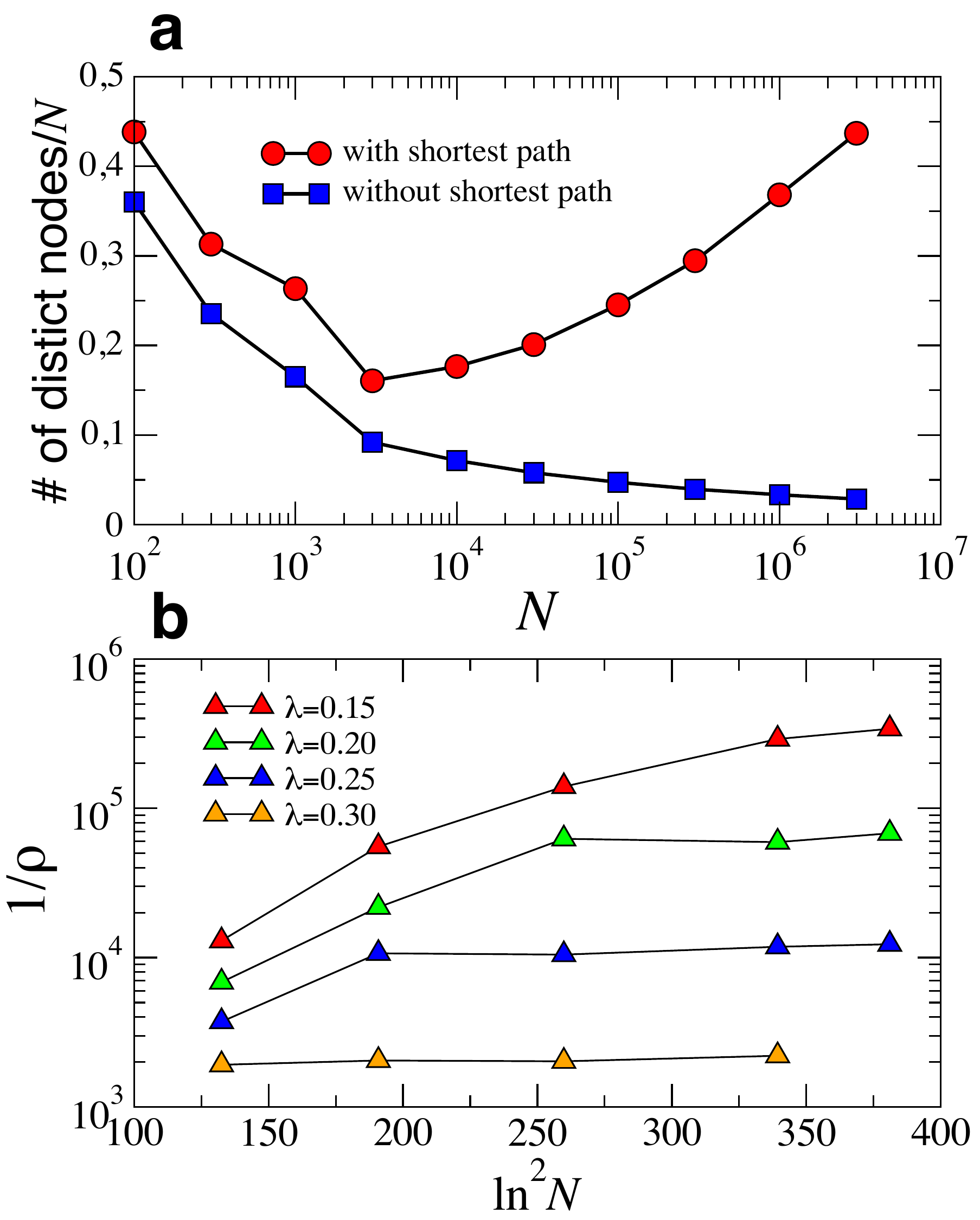}
\caption{{\bf a.} Blue squares show the relative size of the subgraph made of
  nodes with $k>\ln{N}$ plus their direct neighbors as a function of
  $N$. As expected, this is a decreasing function of $N$. Red circles
 show the relative size of the same subgraph plus the number
  of distinct nodes connecting any pair of nodes with $k>\ln{N}$ with
  a shortest path. As it can be observed, this curve increases with
  $N$ (and eventually converges), indicating that this subgraph is
  macroscopic. This happens even if we have only considered one
  shortest path per pair of nodes, so that the red circles are just a
  lower bound of the relative size of the real subgraph. 
  {\bf b.} Behavior of the inverse of the prevalence, as measured from the
  quasi-stationary method, as a function of the system size for a SF
  network with $\gamma=4$ and $k_{min}=2$. 
  Values of the system size $N$ range from $10^5$
to $3\times 10^8$. $\lambda_c^{HMF}=0.328$, $\lambda_{max}(N=10^5)=0.3536$, $\lambda_{max}(N=10^6)=0.2878$,$\lambda_{max}(N=10^7)=0.2248$.}
\label{fig1} 
\end{figure}

\textbf{Bogu\~n\'a \textit{et al.} Reply:} In their comment to our letter~\cite{Boguna:2013fk}, Lee et al.~\cite{Lee:comment} make an important point
concerning the solution of Eq.~(3) in~\cite{Boguna:2013fk}. They are right in pointing out that this
equation predicts, for any given value of $\lambda$, epidemic activity
only for nodes with degrees above $\ln{N}$. From this result, they
conclude that the prevalence $\rho$ is bounded by $(\ln N)^{-(\gamma-2)}$
and, thus, that it goes to zero in the thermodynamic limit. This observation is indeed correct as far as Eq.~(3) is
concerned. However, we would like to point out that Eq.~(3)
does not describe the original dynamics but an effective one that only
considers infections among distant hubs mediated by chains of nodes.
When translated back to the original dynamics in the real network,
this subset of high degree nodes does lead to a truly endemic activity
for any value of $\lambda$.

Indeed, in the modified dynamics, chains of nodes connecting high 
degree nodes are considered only as the way to propagate the infection 
among hubs but not part of the system per se. However, in the real 
dynamics, these nodes do actually belong to the network and, because they are the ones
propagating the epidemics, they are necessarily active. The sum
of nodes with degrees above $\ln{N}$ plus nodes connecting them is, in fact, a 
macroscopic fraction of the entire system, as we show in Fig.~\ref{fig1}~a.
As a consequence, above the threshold $\lambda_{max}(N)$ predicted by the
modified dynamics, the activity in the original system is truly endemic.

To further check our results, we have made more numerical simulations with 
the quasi-stationary method in random networks with $\gamma=4$ and 
$k_{min}=2$. Figure~\ref{fig1}~{\bf b} shows the inverse of the prevalence 
$1/\rho$ in logarithmic scale as a function of
$(\ln{N})^{\gamma-2}$ for values of $\lambda$ below the heterogeneous mean 
field (HMF) prediction $\lambda_c^{HMF}$.
If the argument by Lee et al. were correct, we should find a linear 
increasing function for any value of
$\lambda<\lambda_c^{HMF}$. Instead, we only observe a growing behavior for 
values of $\lambda$ smaller than our prediction $\lambda_{max}(N)$, 
whereas we find a constant value when $\lambda_{max}(N)<\lambda<\lambda_c^{HMF}$.

We therefore conclude that the main result of our manuscript, namely
the bound of the threshold of the SIS model as given implicitly by
Eq.~(4) in Ref.\cite{Boguna:2013fk}, is essentially correct, providing
a reasonable theoretical estimate of its value.

\vspace*{0.25cm}

\small
\noindent
M.~Bogu\~n\'a$^1$, C.~Castellano$^2$, and R.Pastor-Satorras$^3$\\ 
$^1$Departament de F\'{\i}sica Fonamental, Universitat de Barcelona, 08028 Barcelona, Spain\\
$^2$Istituto dei Sistemi Complessi (ISC-CNR), I-00185 Roma, Italy and
Dipartimento di Fisica, ``Sapienza'' Universit\`a di Roma, I-00185 Roma, Italy\\ 
$^3$Departament de F\'\i sica i Enginyeria Nuclear, Universitat
Polit\`ecnica de Catalunya, 08034 Barcelona, Spain

\nobreak


\begin{thebibliography}{2}
\expandafter\ifx\csname natexlab\endcsname\relax\def\natexlab#1{#1}\fi
\expandafter\ifx\csname bibnamefont\endcsname\relax
  \def\bibnamefont#1{#1}\fi
\expandafter\ifx\csname bibfnamefont\endcsname\relax
  \def\bibfnamefont#1{#1}\fi
\expandafter\ifx\csname citenamefont\endcsname\relax
  \def\citenamefont#1{#1}\fi
\expandafter\ifx\csname url\endcsname\relax
  \def\url#1{\texttt{#1}}\fi
\expandafter\ifx\csname urlprefix\endcsname\relax\def\urlprefix{URL }\fi
\providecommand{\bibinfo}[2]{#2}
\providecommand{\eprint}[2][]{\url{#2}}

\bibitem[{\citenamefont{Bogu\~n\'a et~al.}(2013)\citenamefont{Bogu\~n\'a,
  Castellano, and Pastor-Satorras}}]{Boguna:2013fk}
\bibinfo{author}{\bibfnamefont{M.}~\bibnamefont{Bogu\~n\'a}},
  \bibinfo{author}{\bibfnamefont{C.}~\bibnamefont{Castellano}},
  \bibnamefont{and}
  \bibinfo{author}{\bibfnamefont{R.}~\bibnamefont{Pastor-Satorras}},
  \bibinfo{journal}{Phys. Rev. Lett.} \textbf{\bibinfo{volume}{111}},
  \bibinfo{pages}{068701} (\bibinfo{year}{2013}),
  \urlprefix\url{http://link.aps.org/doi/10.1103/PhysRevLett.111.068701}.

\bibitem[{\citenamefont{Lee et~al.}()\citenamefont{Lee, Shim, and
  Noh}}]{Lee:comment}
\bibinfo{author}{\bibfnamefont{H.~K.} \bibnamefont{Lee}},
  \bibinfo{author}{\bibfnamefont{P.-S.} \bibnamefont{Shim}}, \bibnamefont{and}
  \bibinfo{author}{\bibfnamefont{J.~D.} \bibnamefont{Noh}},
  \bibinfo{note}{arXiv:1309.5367}.

\end{thebibliography}
\end{document}